# A FLOSS Tool for Antenna Radiation Patterns

*Nikolitsa YANNOPOULOU[1], Petros ZIMOURTOPOULOS[2]*

[1]Antennas Research Group, Palaia Morsini, Xanthi, Thrace, Hellas, EU
[2]Antennas Research Group, Department of Electrical Engineering and Computer Engineering, Democritus University of Thrace, V. Sofias 12, 671 00 Xanthi, Greece

yin@antennas.gr, pez@antennas.gr, www.antennas.gr

**Abstract.** *This paper briefly highlights the features of the software tool [RadPat4W], named after Radiation Patterns for Windows, that is based on an alternative exposition of fundamental Antenna Theory. This stand-alone application is compatible with the [Wine] environment of Linux and is part of a freeware suite, which is under active development for many years. Nevertheless, the [RadPat4W] source code has been now released as FLOSS Free Libre Open Source Software and thus it may be freely used, copied, modified or redistributed, individually or cooperatively, by the interested user to suit her/his personal needs for reliable antenna applications, from the simplest to the more complex.*

## 1. Introduction

Useful software has to work exactly as someone wants, so the authors' group decided to develop its own mini-Suite of software tools for antenna applications [1]. This project started in the middle of 90s, when the PC with WWW access became power enough to cover the increased requirements of antenna analysis and design, as well as of their results presentation and distribution. Since then, the development of the mini-Suite has been orientated towards the personal needs of the individual user or of the independent member of a small, open, loosely connected group, like the authors' one, who is interested in antenna education, research and engineering, i.e., a student, an educator, a researcher, a professional engineer or a radio amateur. Such a user has enough bibliographical resources provided by the Open Access movement, but only limited technical resources for construction and measurement. The mini-Suite is intended then for the informed user who at least can construct an experimental thin-wire antenna model and at most has access to a VNA Vector Network Analyzer to test this model -by the way, nowadays, the cost of a certified refurbished VNA is just a small percentage of its new price. For that reason, the mini-Suite specifically includes the stand-alone application [RadPat4W].

## 2. [RadPat4W]

The active development of this tool attempts to bridge the increasing gap of today approximate simulation techniques, which dominate antenna applications, to classic exact analysis methods, which concern the demanding user who wants to know what s/he is really doing with these marvellous antenna simulators. To achieve this goal and facilitate the study of antenna application results, either approximated or exact, [RadPat4W] computes and/or plots the antenna geometry, its characteristics, as well as 2D main-plane cuts of its radiation pattern and 3D Virtual Reality objects for its geometry and pattern. Currently, the tool uses by default: (1) working formulas produced by the analysis method of the authors' alternative exposition of fundamental Antenna Theory [2] that is quickly but rigorously results in the most general complex vector expression for the radiation pattern of any thin-wire antenna, and (2) numerical results from approximation techniques based on the Moment Method implemented by the two antenna simulators [DA] and [RichWire], which are included in the mini-Suite [1]. Finally, to support the serious user to judge the results, the current beta version of [RadPat4W] incorporates the superposition on the plotted results of scientific VNA measurements with systematic errors first estimated by the authors in 2008 [3], a process that is now accomplished semi-manually using a combination of other separate mini-Suite tools.

Besides [RadPat4W], which was always distributed through the internet as freeware, other non-commercial software, less related to [RadPat4W] and from developers with a diverse knowledge of Antenna Theory, is distributed under various terms of use. These ware kinds of the free have been exhaustively examined by the members of the USENET group [alt.comp.freeware] with the purpose to warn the candidate user about the actual content of the corresponding licenses. However, to the best of authors' knowledge, it seems that there is still no completely FLOSS for reliable antenna applications. Therefore, with the aim to further encourage the independent user to tweak [RadPat4W] according to her/his personal needs or even to be involved in this modern, most promising, cooperative activity of the FLOSS movement, the authors decided lately to release the entire source code of [RadPat4W] under the approved, by the OSI Open Source Initiative, MIT License.

The source code, now in version 4.4 with help in version 1.0, is developed from scratch, without using any other code, in MS Visual Basic 6 SP6 for 32-bit MS Windows and the executable, which is also compatible with the [Wine] environment of Linux, needs about 8.5 MB of free hard disk space for its installation. The application usability has been multiple checked during antenna courses and theses elaboration, thus its source code is in a mature state for a long time now although, from time to time, new features are added to it. The code is available for download from authors' group website http://www.antennas.gr/floss or GoogleCode repository at http://code.google.com/p/rga/.

The features of [RadPat4W] can be divided according to their functionality in two groups: (1) pattern computation and plotting using working formulas from Antenna Theory, and (2) pattern plotting using numerical results from antenna simulators. To exemplify these features by examples, a number of antenna education, research and engineering applications, from the simplest to the more complex, are presented in the following.

## 3. Working Formulas

Fig. 1 shows the software application form of [RadPat4W] for the three main-plane cuts of the E-normalized radiation pattern of a linear, center-fed, standing-wave dipole, which is parallel to z-axis and has a length of 2.35λ, where the length is the only one input parameter with values in the range [0.001, 10]λ. To overcome the practical constrains of the limited number of screen pixels that obscures the detailed view of zero-E directions, which determine the radiation pattern lobes, a useful feature has been introduced in all application forms that is the magnification of the pattern up to six times. Each magnified diagram shows the zero-E directions by magenta colored radial lines on the screen. The [MaxZero-D] button opens the window of Fig. 2 in which the computed directions of maxima and zeros, as well as, the directivity and the maximum value of pre-normalized E radiation pattern, are shown [2].

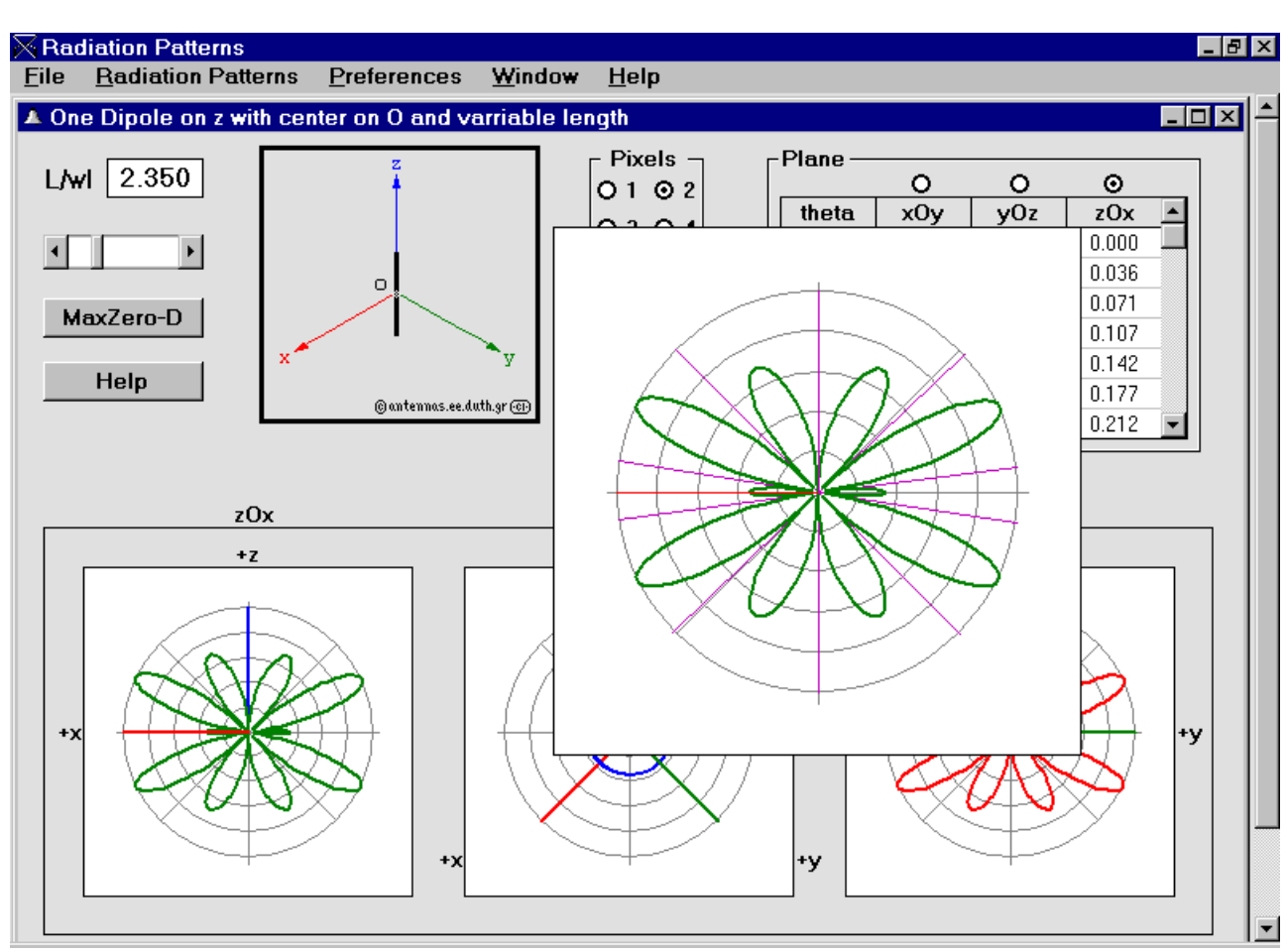

**Fig. 1.** [One Dipole on z]: A magnified main–plane cut.

Max And Zero values of Radiation Pattern - Directivity

| # Zero | theta |
|---|---|
| 1 | 0.00 |
| 2 | 45.40 |
| 3 | 81.43 |
| 4 | 98.57 |
| 5 | 134.60 |
| 6 | 180.00 |

| theta | Max |
|---|---|
| 64 | 1.61 |
| 116 | 1.61 |
| Directivity | 2.83 |
| dBi | 4.52 |
| dBd | 2.37 |

Back

**Fig. 2.** [One Dipole on z]: Maximum, zeros and Directivity.

Fig. 3 shows the three main-plane cuts of radiation pattern for the same dipole but in the space direction described by the input data of its unit directional vector: (0.000, 0.707, –0.707). The [Zero] button opens the window of Fig. 4, where the computed directions of zeros on the three main-plane cuts, as well as on a plane that contains the dipole axis, are shown.

**Fig. 3.** [One Dipole in Space]: Three different main–plane cuts.

Zero values of Radiation Pattern

| # Zero | $\xi_\ell$ |
|---|---|
| 1 | 0.00 |
| 2 | 45.40 |
| 3 | 81.43 |
| 4 | 98.57 |
| 5 | 134.60 |
| 6 | 180.00 |

| xOy | | yOz | | zOx | |
|---|---|---|---|---|---|
| th90 | | ph90 | ph270 | ph0 | ph180 |
| 12.16 | 263.27 | .41 | 44.99 | 6.88 | |
| 83.27 | 276.73 | 36.44 | 90.39 | 77.84 | |
| 96.73 | 347.84 | 53.57 | 126.43 | 102.16 | |
| 167.84 | | 89.61 | 143.56 | 173.12 | |
| 192.16 | | 135.01 | 179.59 | | |

Back

**Fig. 4.** [One Dipole in Space]: Zero pattern directions.

Fig. 5 shows the input data for a uniform linear array of exactly parallel linear standing-wave dipoles, in two frames for the generator dipole with complex vector pattern G and the array of isotropic sources with complex number pattern A, respectively. The first frame defines the length and the direction of the generator or reference dipole. The second frame defines the geometrical and electrical input data for the array: the number of isotropic point sources, their constant phase difference in degrees, their constant equidistance per wavelength, and the unit directional vector of array axis. The shown values are for an array of 3 dipoles, 2λ each, in the direction of +y axis, with phase difference of −144°, 0.4λ apart, on the array axis direction of +x. The [Show Graphics] button opens the window of Fig. 6, in which the dipole array complex vector pattern E = AG, i.e., the Principle of Radiation Patterns Multiplication, is shown in absolute 3D form: the normalized norm pattern $\|E\|$ results as product of the normalized absolute pattern $|A|$ by the normalized norm pattern $\|G\|$ and by a non-shown constant spherical pattern $|A|_{max}\|G\|_{max}/\|E\|_{max} \geq 1$ [2]. In the [Directivity] frame of the window, the directivities of the array $D_A$, of the generator $D_G$, and of the dipole array D are shown. The [Max-Zero of Array Factor] button opens the window of Fig. 7, in which the directions of zero-A and $|A|_{max}$ and the values of $|A|_{max}$, $\|G\|_{max}$ and $\|E\|_{max}$ are shown.

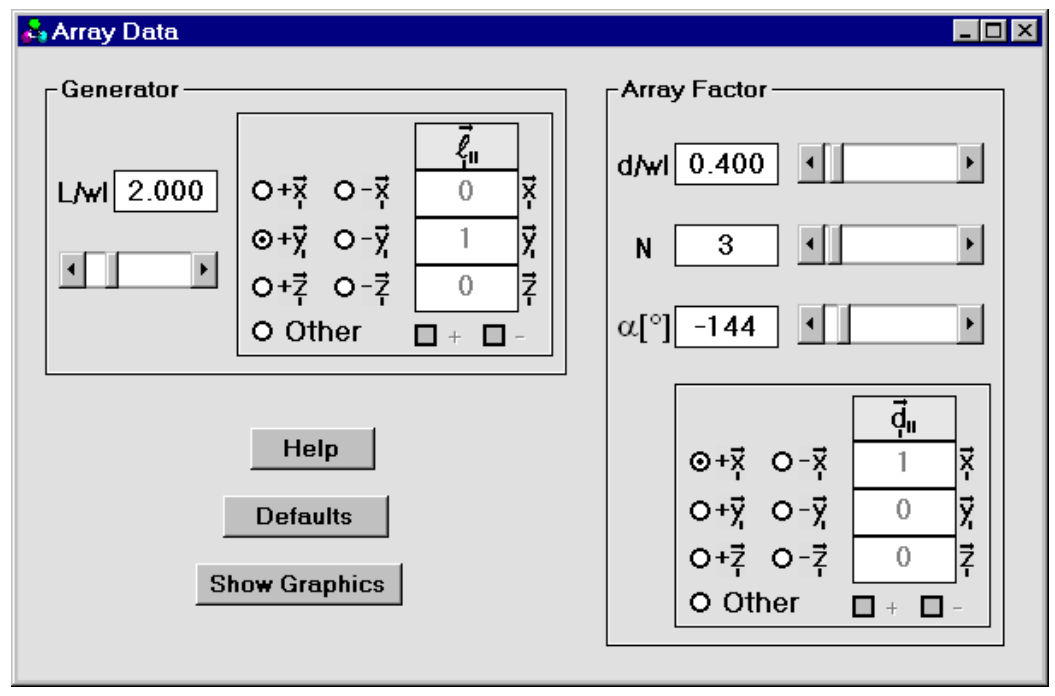

**Fig. 5.** [Array Data]: Uniform Linear Array input.

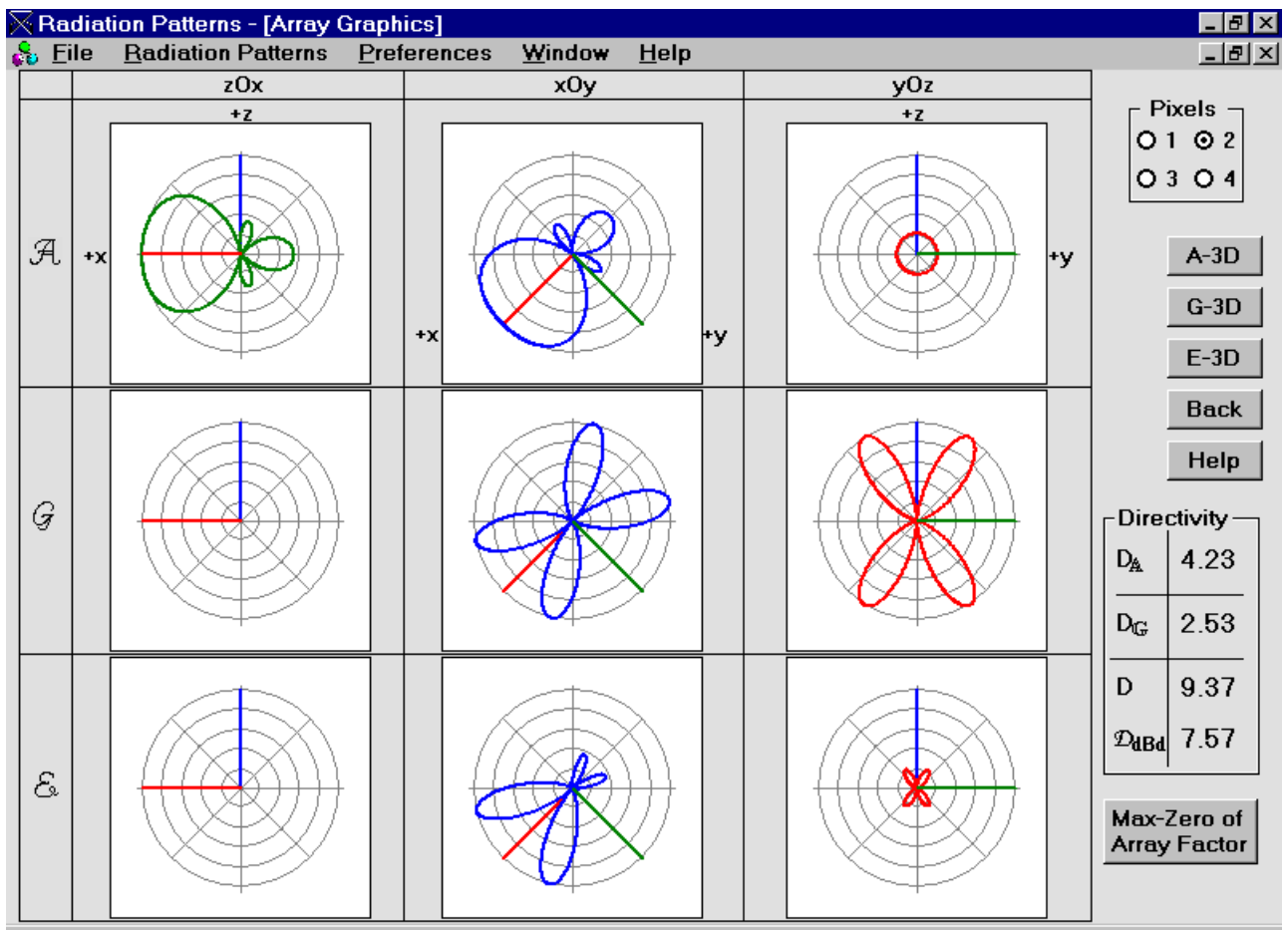

**Fig. 6.** [Array Graphics]: 9 main–plane pattern cuts with 2 zeros.

Max and Zero Values of Array Factor - Max Values of Patterns

| # Zero | $\xi_d$ |
|---|---|
| 1 | 80.41 |
| 2 | 131.81 |

| # Max | $\xi_d$ |
|---|---|
| 1 | 0.00 |

Max of Patterns

| | | |
|---|---|---|
| Emax 2D = **6.69** | Amax 2D = **3.00** | Gmax 2D = **2.34** |
| Emax 3D = **6.67** | Amax 3D = **3.00** | Gmax 3D = **2.34** |

Back

**Fig. 7.** [Max-Zero of Array Factor]: Computed results.

The buttons [A-3D], [G-3D] and [E-3D] produce the respective 3D Virtual Reality radiation patterns, which are shown in Fig. 8 as three screen captures of the free Platinum WorldView VRML viewer plug-in for MS Internet Explorer. By the way, the contemporary free VRML Cortona viewer is available for a number of other web browsers too, under MS Windows, while under Linux the most appropriate add-on for iceweasel (Mozilla Firefox) is the FreeWRL VRML viewer.

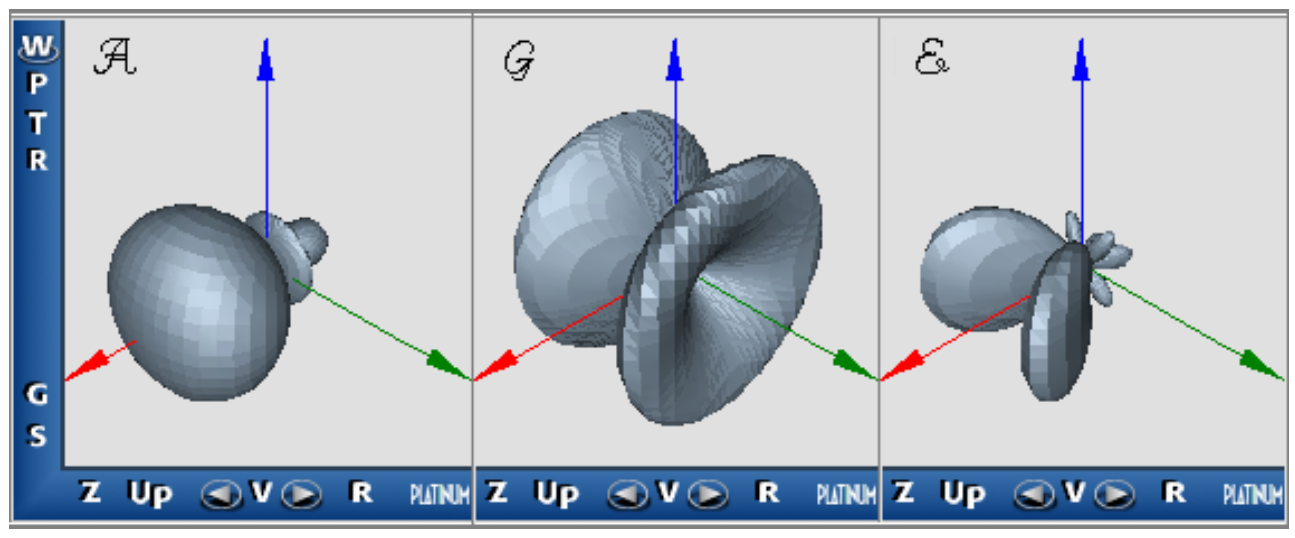

**Fig. 8.** Virtual 3D Principle of Radiation Pattern Multiplication.

## 4. Antenna Simulators

The second group of [RadPat4W] features concerns its ability to plot antenna geometry and patterns from the clear text data files such those of the two antenna simulators of the mini-Suite: [DA] and [RichWire]. The source code of [DA], now in version 1.0.8, has been written in Compaq Visual Fortran 6.1 as Quick-Win 32-bit application for MS Windows, and the executable, which is also compatible with the [Wine] environment of Linux, can be easily installed on a PC with free hard disk space of only about ~500 KB. In essence, this application is a restricted variation of [RichWire], which is a fully analyzed, corrected and redeveloped edition of the original Moment Method thin-wire computer program by J.H.Richmond, available in the public domain by NASA since 2005 [4]. [DA] is used for antenna simulation by half-wave dipoles, with just one active. The program requires an input data file to derive three output data files. All these data files are used by [RadPat4W]. In Fig. 9, the simplest data for only one dipole in space are shown. [RichWire] data files are similar. The usability of both antenna simulators has been also multiple checked. The simulators are available as freeware from the mentioned repositories.

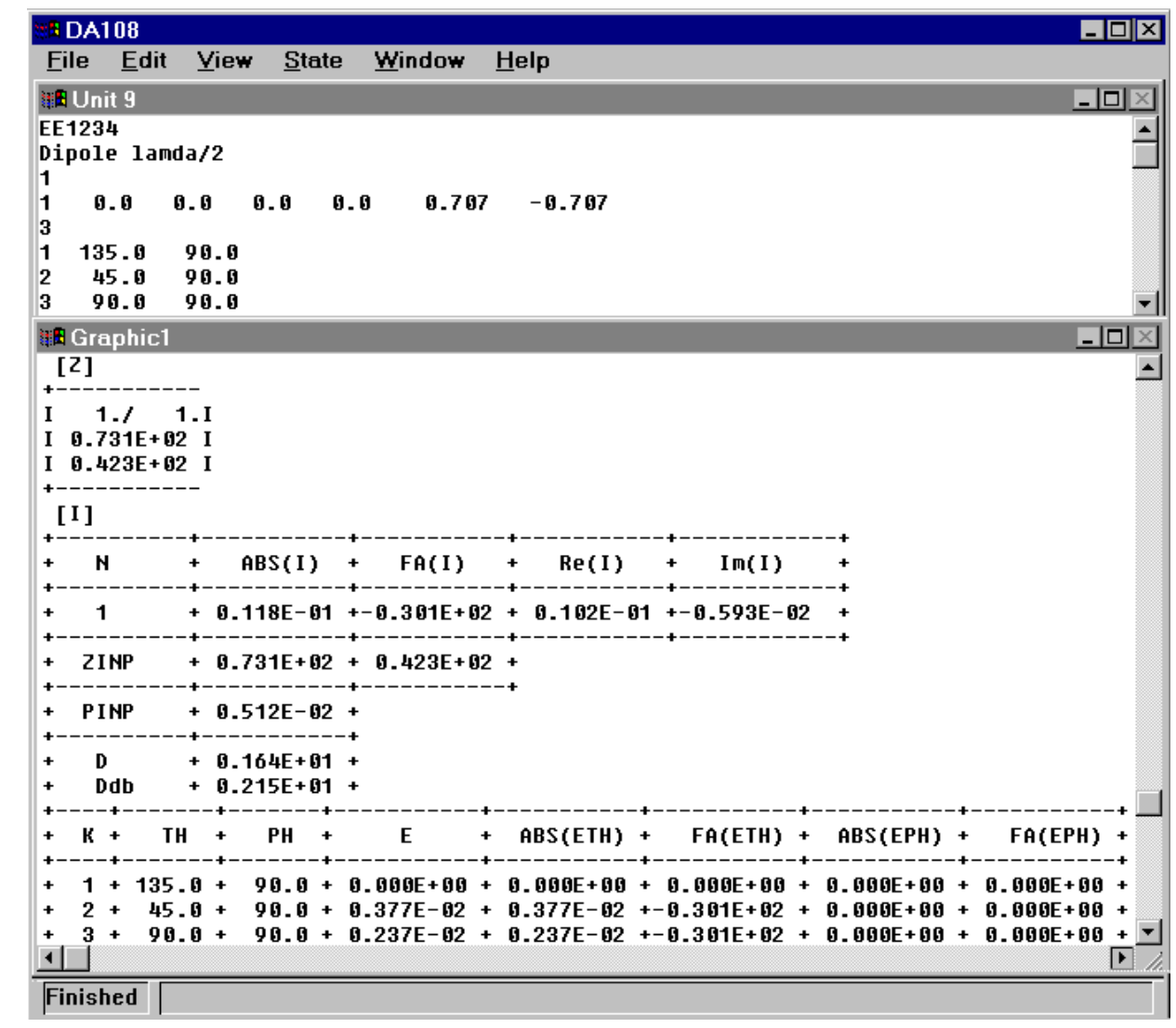

**Fig. 9.** [Da]: Input and output data for the simplest case.

In Fig. 10, the [RadPat4W] engineering application for a commercial VHF Yagi-Uda antenna is shown. The [Antenna] button reads the antenna geometry [RichWire] data input file. The [geo.wrl] button produces the 3D Virtual Reality antenna geometry and simultaneously opens the [GL Viewer for Mathematica] for an immediate view [5]. In Fig. 11, the drawn results produced by [RadPat4W] are shown for an educational application of a flat airplane modeled with non-overlapping $\lambda/2$ dipoles in [DA] [6]. Fig. 12 illustrates the drawn results produced by the currently beta version of [RadPat4W] for a research application of a constructed Hentenna model, simulation designed with [RichWire] and measured with a VNA system [7].

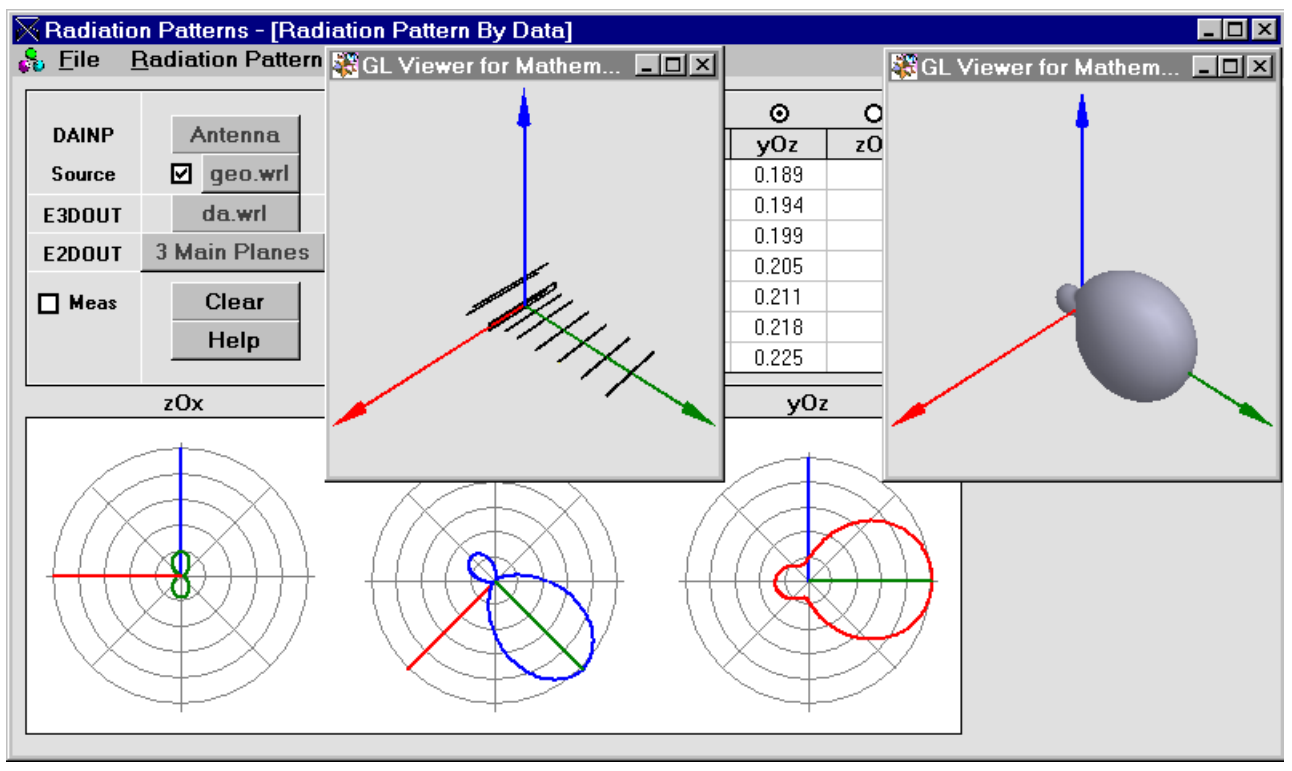

**Fig. 10.** [RichWire]: A commercial VHF Yagi–Uda antenna.

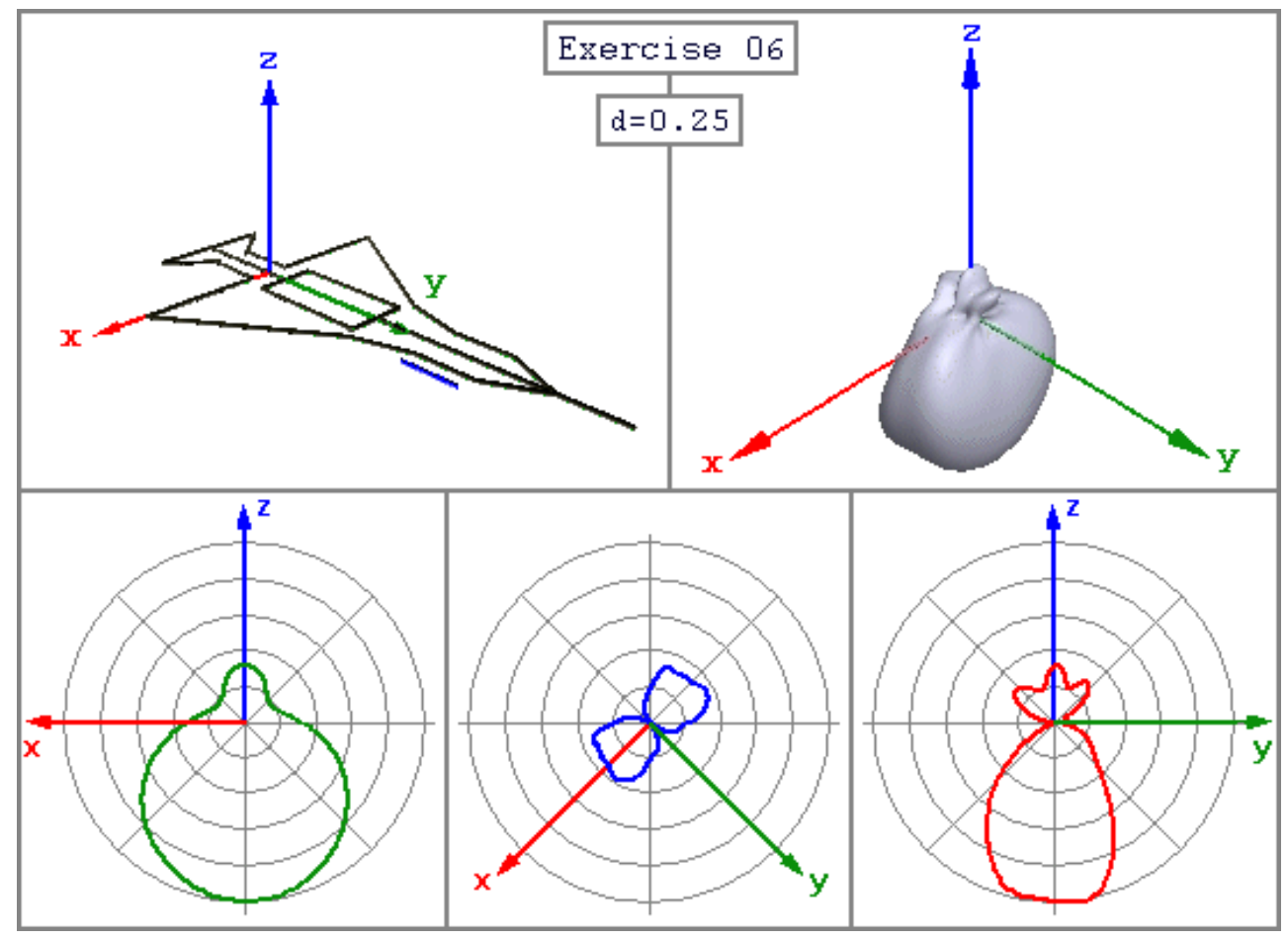

**Fig. 11.** [RadPat4W]: Results for a flat airplane model.

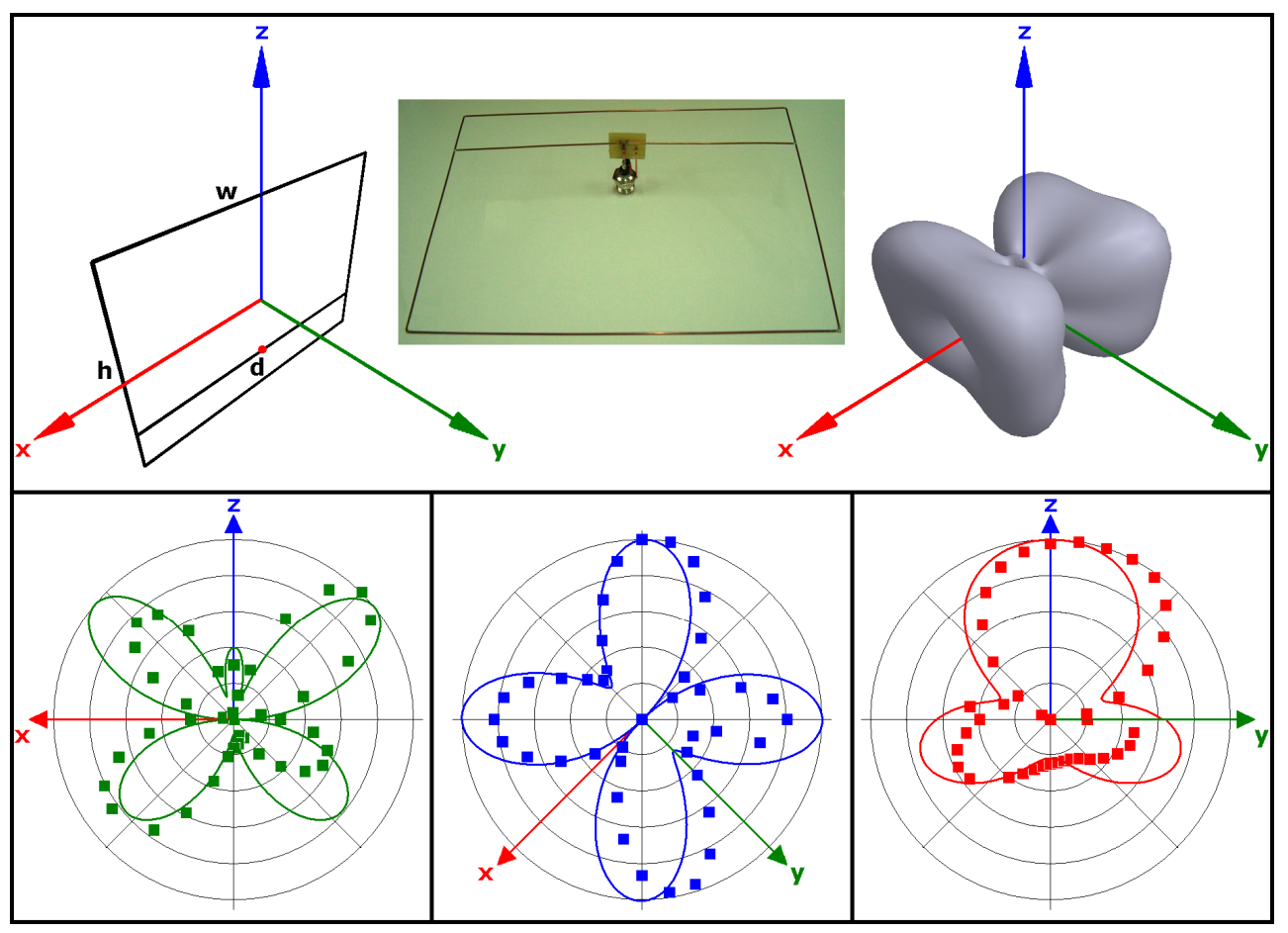

**Fig. 12.** [RadPat4W]: Design, Construction and Measurement.

## 5. [RadPat4W] Development Plans

Scheduled expansions of [RadPat4W] include the following facilities, which are already available in other mini-Suite tools: (1) choice of other plane- or conical-cuts, (2) key-in of any exact analysis working formula $E(\theta,\phi)$, (3) selection of the $\xi$-, $\theta$-, and $\phi$- plotting step, (4) automation of fine Cartesian pattern plotting, (5) superposition of VNA measurements on 3D Virtual Reality patterns, such as that in the left part of Fig. 13 [8], and (6) superposition of VNA measurements with their differential error cloud [3] on 2D plots, such that in the right part of Fig. 13 [9].

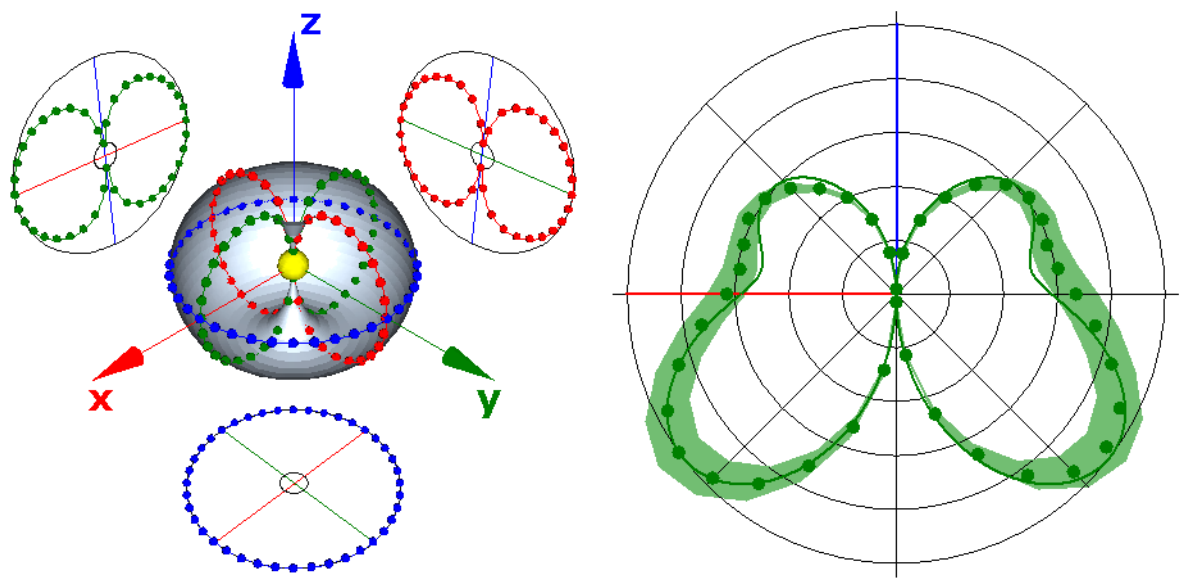

**Fig. 13.** 3D VNA measurements and their 2D error cloud.

Any other expansion of the freely available code is of course welcomed. In authors' group, there are no plans in the near future to upgrade the mini-Suite to an integrated environment for its tools.